\documentclass{article}

\usepackage{arxiv}

\usepackage[utf8]{inputenc} \usepackage[T1]{fontenc}    

\PassOptionsToPackage{hyphens}{url}

\usepackage[breaklinks,backref=page]{hyperref}

\usepackage{url}            \usepackage{booktabs}       \usepackage{amsfonts}       \usepackage{nicefrac}       \usepackage{microtype}      \usepackage{lipsum}         \usepackage{graphicx}
\usepackage[numbers,sort&compress]{natbib}
\usepackage{doi}
\usepackage{float}
\usepackage{multirow}
\usepackage{lineno}

\usepackage{cleveref}

\usepackage{wrapfig}

\usepackage{authblk}

\usepackage[justification=centering]{caption}

\usepackage[inline]{enumitem}

\usepackage{bbding}

\usepackage[flushleft]{threeparttable}

\usepackage{xcolor}

\usepackage{listings}

\usepackage{algorithm}
\usepackage[noend]{algpseudocode}
\usepackage{algorithmicx}

\usepackage{etoolbox}

\newcommand{\algstrut}[1][\algruledefaultfactor]{\vrule width 0pt
depth .25\baselineskip height #1\baselineskip\relax}
\newcommand*{\algrule}[1][\algorithmicindent]{\hspace*{.5em}\vrule\algstrut[.95]
\hspace*{\dimexpr#1-.5em}}

\makeatletter
\newcount\ALG@printindent@tempcnta
\def\ALG@printindent{\ifnum \theALG@nested>0\ifx\ALG@text\ALG@x@notext \else
    \unskip
\ALG@printindent@tempcnta=1
    \loop
    \algrule[\csname ALG@ind@\the\ALG@printindent@tempcnta\endcsname]\advance \ALG@printindent@tempcnta 1
    \ifnum \ALG@printindent@tempcnta<\numexpr\theALG@nested+1\relax \repeat
    \fi
    \fi
}

\patchcmd{\ALG@doentity}{\noindent\hskip\ALG@tlm}{\ALG@printindent}{}{\errmessage{failed to patch}}

\AtBeginEnvironment{algorithmic}{\lineskip0pt}

\renewcommand*{\backref}[1]{}
\renewcommand*{\backrefalt}[4]{
  \ifcase #1
    No citations.\or
(Cited on page: #4).
  \else
(Cited on pages: #4).
  \fi
}

\title{SC2Tools: StarCraft II Toolset and Dataset API }

\date{May 5, 2025}

\makeatletter
\newcommand\email[2][]{\newaffiltrue\let\AB@blk@and\AB@pand
      \if\relax#1\relax\def\AB@note{\AB@thenote}\else\def\AB@note{\relax}
        \setcounter{Maxaffil}{0}\fi
      \begingroup
        \let\protect\@unexpandable@protect
        \def\thanks{\protect\thanks}\def\footnote{\protect\footnote}
        \@temptokena=\expandafter{\AB@authors}
        {\def\\{\protect\\\protect\Affilfont}\xdef\AB@temp{#2}}
         \xdef\AB@authors{\the\@temptokena\AB@las\AB@au@str
         \protect\\[\affilsep]\protect\Affilfont\AB@temp}
         \gdef\AB@las{}\gdef\AB@au@str{}
        {\def\\{, \ignorespaces}\xdef\AB@temp{#2}}
        \@temptokena=\expandafter{\AB@affillist}
        \xdef\AB@affillist{\the\@temptokena \AB@affilsep
          \AB@affilnote{}\protect\Affilfont\AB@temp}
      \endgroup
       \let\AB@affilsep\AB@affilsepx
}
\makeatother

\author[1]{\textbf{Andrzej Białecki}\textsuperscript{*,}} 

\author[2]{\textbf{Piotr Białecki}}

\author[1,4]{\textbf{Piotr Sowiński}}

\author[2]{\textbf{Mateusz Budziak}}

\author[3]{\textbf{Jan Gajewski}}

\affil[1]{Warsaw University of Technology}
\affil[2]{Independent Researcher}
\affil[3]{Józef Piłsudski University of Physical Education in Warsaw}
\affil[4]{NeverBlink, Poland}

\hypersetup{
pdftitle={SC2Tools: StarCraft II Toolset and Dataset API},
pdfauthor={Andrzej Białecki},
pdfkeywords={gaming, esport, data processing, toolset, dataset preparation},
}

\begin{document}
\maketitle

\let\thefootnote\relax\footnote{\textsuperscript{*} Corresponding author: \url{andrzej.bialecki94@gmail.com}}
\let\thefootnote\relax\footnote{\textsuperscript{*} Institutional contact: \url{andrzej.bialecki.dokt@pw.edu.pl}}

\begin{abstract}

    Computer games, as fully controlled simulated environments, have been utilized in significant scientific studies demonstrating the application of Reinforcement Learning (RL). Gaming and esports are key areas influenced by the application of Artificial Intelligence (AI) and Machine Learning (ML) solutions at scale. Tooling simplifies scientific workloads and is essential for developing the gaming and esports research area.

In this work, we present ``SC2Tools'', a toolset containing multiple submodules responsible for working with, and producing larger datasets. We provide a modular structure of the implemented tooling, leaving room for future extensions where needed. Additionally, some of the tools are not StarCraft~2 exclusive and can be used with other types of data for dataset creation.

The tools we present were leveraged in creating one of the largest StarCraft~2 tournament datasets to date with a separate PyTorch and PyTorch Lightning application programming interface (API) for easy access to the data.

We conclude that alleviating the burden of data collection, preprocessing, and custom code development is essential for less technically proficient researchers to engage in the growing gaming and esports research area. Finally, our solution provides some foundational work toward normalizing experiment workflow in StarCraft~2 
\end{abstract}


\section{Introduction and Background}
\label{sec:MotivationAndSignificance}

Computer games as fully controlled simulated environments were used in major scientific works that showcased the application of Reinforcement Learning (RL). As such, computer games can be viewed as one of the many components of major breakthroughs and advancements in RL applications \cite{Szita2012RLGames,Samsuden2019RLGames,LanctotEtAl2019OpenSpiel,Shao2019RLSurvey,Jayaramireddy2023RLSurvey,Vinyals2019,Wurman2022}.

Despite heightened interest in research on gaming and esports, there are limited high-level libraries and tools made for rapid experimentation in some game titles. Researchers from various research disciplines have shown their interest in exploring gaming and esports, including: (1) psychology \cite{Campbell2018}, (2) computer science \cite{Rashid2020,Yuan2021ActorCritic}, (3) education \cite{Jensen2024,Jenny2021}, (4) medical sciences \cite{Krarup2020109344}, and others \cite{Holden2017Law,Nagorsky2020}.
The ability to tie these topics with the in-game data cannot be overstated.

When such software is available, it is often hard to use for less technically proficient researchers. Data parsing libraries are prevalent in computer games, such as Counter-Strike \cite{AWPYXeno2020,ClarityGitHub}, Rocket~League \cite{URLBoxcars2016}, Dota~2 \cite{OpenDotaGitHub,ClarityGitHub}, and finally in StarCraft~2 \cite{URLBlizzardS2ClientProto,URLS2Prot2016,GitHubSC2Reader}.

Esports can be treated as a subset of gaming with additional requirements for players, such as tournament presence, organized play, training, and professionalization \cite{Formosa2022}. The study of esports is multidisciplinary in nature \cite{Brock2023Interdisciplinary,Pizzo2022Interdisciplinary}. Due to the growing academic interest in the area of gaming and esports \cite{Yamanaka2021Review,Reitman2020Review,Tang2023Review,Bialecki2024Review}, it is key to provide tools for researchers capable of simplifying the process of acquiring large datasets efficiently, not only for authors interested in the area of computer science \cite{Ferenczi2024sc2_serizlizer,SmerdovLoLDataset2021}.

In case of our implementation, we focus on solving problems within the StarCraft~2 (SC2) infrastructure ecosystem. StarCraft~2 is a real-time strategy game developed by Blizzard Entertainment. The game is known as one of the most prominent real-time strategy (RTS) esports titles \cite{Tyreal2020,Dal2020}. It is also characterized by its fast-paced gameplay and a high skill ceiling \cite{Migliore2021}. These attributes make for a great environment for testing various AI agents \cite{Ma2024LLMStarCraft2,Vinyals2019,Samvelyan2019SMAC,Pearce2022CSGO}. Moreover, research in StarCraft~2 is not limited to AI agents -- there are efforts to analyze the game from various perspectives and provide insights that can assist players in their gameplay \cite{URLSC2AICoach2022LLM,URLSc2replaystats}.

Our software collection is an open-source implementation of data extraction, and data interfacing tools for StarCraft~2. We solve the problem of ease of access to the data encoded in files with ".SC2Replay" extension by using an open-source file extractor for proprietary MoPAQ (MPQ) file format. From this point on, we will refer to the MPQ files with the ".SC2Replay" extension as SC2Replay files.

So far, our software was leveraged in preparation of major datasets: "SC2ReSet" \cite{Bialecki2022ReSetZenodo} and "SC2EGSet" \cite{Bialecki2023EGSetZenodo} with an accompanying peer-reviewed and published Data Descriptor article \cite{Bialecki2023SC2EGSet}. The output of our software was used in varying contexts indirectly. authors cited our work, some of them following the general flow of our exploration \cite{Kim2024}. Others put emphasis on statistical calculation within esports landscape \cite{Dupuy2024}. Finally authors describe our work in surveys of related work when working in another games \cite{Johar2024}.

Our solution uses the official StarCraft 2 replay file format specification provided via Blizzard Entertainment GitHub repository \cite{URLBlizzards2protocol2013}. Specifically, in "SC2InfoExtractorGo", we extend the community-built Golang implementation of the parser \cite{URLS2Prot2016,URLMPQ2017}. The output of our software pipeline is a fully prepared dataset, ready for use with our extension of PyTorch \cite{PyTorch2019} and PyTorch Lightning interfaces \cite{PyTorch_Lightning_2019}. Our goal was to lower the technical knowledge required to obtain data from in-game replays.

\section{Software Description}
\label{sec:SoftwareDescription}

Our software publication consists of multiple modules that the user can match to their specific needs. To easily extend our toolset, the main repository of ``SC2Tools'' contains multiple git submodules. Each submodule is a separate repository with the logic required to perform a specific tasks on the SC2Replay files. The motivation for this structure is twofold. Firstly, it makes evolving the toolset easier, as modules can be easily replaced, or new ones added. Secondly, users have the option of using only a portion of the pipeline. The full list of current submodules is as follows: (1) ``SC2InfoExtractorGo'' \cite{Bialecki_2021_SC2InfoExtractorGo}, (2) ``DatasetPreparator'' \cite{Bialecki_2022_SC2DatasetPreparator}, (3) ``SC2AnonServerPy'' \cite{Bialecki_2021_SC2AnonServerPy}. (4) ``SC2\_Datasets'' \cite{bialecki_2022_sc2datasets}.

In case of developing future tools, deprecating or improving the existing implementations, updates will be made to the common open-source repository. In their current state, all of the tools can be either used independently or combined in a data processing pipeline. The pipeline can interoperate with other software tools, as long as they use standard SC2Replay files for inputs to the ``SC2InfoExtractorGo''. Finally, loading the data for experiments is supported as long as the output is saved in Java Script Object Notation (JSON) files with ".json" extension, and conforms to a pre-defined schema defined by the ``SC2\_Datasets'' parser. From now on, we will refer to such files as JSON files. In the current version extending the software with additional tools is possible, and we encourage the community to contribute to the project for future releases.

\subsection{Software Architecture}
\label{sec:SoftwareArchitecture}

\begin{figure}[H]
    \includegraphics[width=\linewidth]{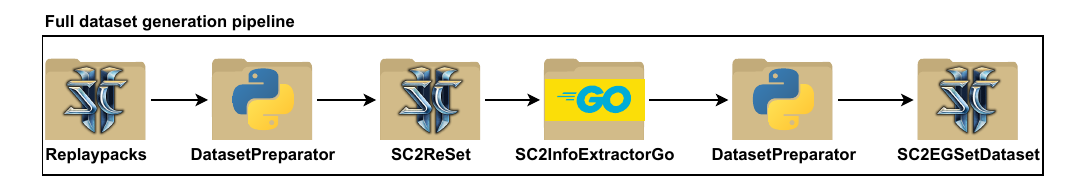}
    \caption{Simplified full pipeline using SC2Tools to create two datasets, ``SC2ReSet'' \cite{Bialecki2022ReSetZenodo} and ``SC2EGSet'' Dataset \cite{Bialecki2023EGSetZenodo}. Initially introduced in \cite{Bialecki2023SC2EGSet}. \autoref{fig:DatasetPipeline}}
    \label{fig:DatasetPipeline}
\end{figure}

Tools and scripts in our repository have singular responsibilities. Each of our submodules fulfills a specific part of the data processing needs within the pipeline. The full pipeline in a simplified pictorial form is showcased in \autoref{fig:DatasetPipeline}. Note the distinctive steps of data pre-processing are explained in \nameref{sec:DatasetPreparator} introducing the ``DatasetPreparator'' Python utility scripts. Further, data processing Golang tool implementatino is explained in \nameref{sec:SC2InfoExtractorGo} introducing the ``SC2InfoExtractorGo''. Finally, data modeling or post-processing tasks are introduced in \nameref{sec:SC2Datasets}  diving deeper on "SC2\_Datasets" as a Python API implementations for PyTorch \cite{PyTorch2019} and PyTorch Lightning \cite{PyTorch_Lightning_2019}.

\subsubsection{DatasetPreparator}
\label{sec:DatasetPreparator}

The ``DatasetPreparator'' \cite{Bialecki_2022_SC2DatasetPreparator} submodule is a set of scripts that ease the process of working with major collections of raw data (replaypacks/datasets). A full list of scripts is as follows:
\begin{enumerate} \item ``directory\_flattener.py'';~flattens the nested directory structure of the replaypacks,
    \item ``directory\_packager.py'';~packages all of the directories in the specified input directory,
    \item ``file\_renamer.py'';~renames the files in the directory to follow a specific naming convention (e.g., to match the dataset schema),
    \item ``json\_merger.py'';~merges two JSON files into one,
    \item ``processed\_mapping\_copier.py'';~copies the auxiliary files generated by ``directory\_flattener.py'' to matching output directories. Built specifically to prepare the ``SC2EGSet'' prior to packaging,
    \item ``sc2\_map\_downloader.py'';~wraps ``SC2InfoExtractorGo'' to run the map downloading step,
    \item ``sc2egset\_replaypack\_processor.py'';~wraps ``SC2InfoExtractorGo'' to run the replaypack processing step on multiple directories at once,
    \item ``sc2egset\_pipeline.py'';~wraps the entire processing pipeline used to obtain the ``SC2ReSet'' and ``SC2EGSet'' datasets,
    \item ``sc2reset\_replaypack\_downloader.py'';~downloads the raw (flattened) replaypacks of ``SC2ReSet'' \cite{Bialecki2022ReSetZenodo} for users that wish to use their own tools for data processing.
\end{enumerate}

In the context of our work, this submodule is responsible for preparing directory structure, execution of "SC2InfoExtractorGo" on the data, and packaging the dataset for hosting. Finally, the current capabilities include downloading the raw replaypacks of "SC2ReSet" \cite{Bialecki2022ReSetZenodo} for ease of "SC2EGSet" Dataset reproduction \cite{Bialecki2023EGSetZenodo}.

\subsubsection{SC2InfoExtractorGo}
\label{sec:SC2InfoExtractorGo}

\begin{figure}[H]
    \centering
    \includegraphics[width=\linewidth]{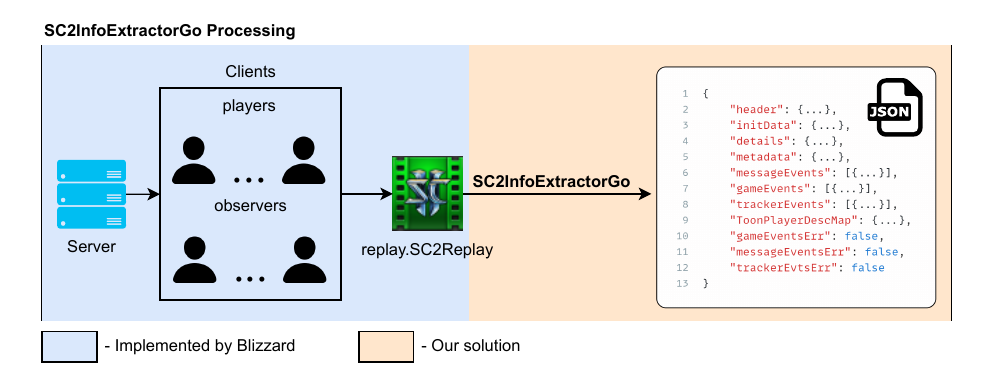}
    \caption{Pictorial representation of the ``SC2InfoExtractorGo'' functionality \cite{Bialecki_2021_SC2InfoExtractorGo}. Replays contain the events which happened during gameplay (blue background), our implementations extracts this data and outputs it for further analysis by the user (orange background).}
    \label{fig:file_processing_sc2infoextractorgo}
\end{figure}

The SC2InfoExtractorGo as a submodule is a tool responsible for extracting the data from SC2Replay files, it depends on previously published open-source lower-level libraries \cite{URLS2Prot2016,URLMPQ2017}. The tool is written in Golang and is shipped as a binary file (release), and as a Docker image via DockerHub. A simplified depiction of the data extraction is available on \autoref{fig:file_processing_sc2infoextractorgo}.

\subsubsection{SC2\_Datasets}
\label{sec:SC2Datasets}

\begin{figure}[H]
    \centering
    \includegraphics[width=0.8\linewidth]{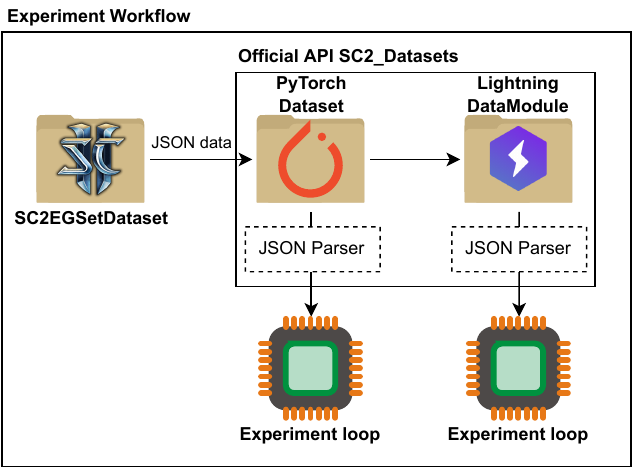}
    \caption{Loading the output of the \nameref{sec:SC2InfoExtractorGo} for machine learning and artificial intelligence use with ``SC2\_Datasets''.}
    \label{fig:LoadingDataToPyTorch}
\end{figure}

One of our solutions, ``SC2\_Datasets'' \cite{bialecki_2022_sc2datasets} interfaces with the JSON files produced by the ``SC2InfoExtractorGo'' \cite{Bialecki_2021_SC2InfoExtractorGo}. This includes all of the classes and methods required to load a single JSON, a collection of JSON files (representing a replaypack), and finally a way of loading an entire dataset (a collection of replaypacks). The pictorial representation of the ``SC2\_Datasets'' functionality is presented on \autoref{fig:LoadingDataToPyTorch}.

Users have the ability to extend our solution and apply it to their data via the PyTorch \cite{PyTorch2019} and PyTorch Lightning \cite{PyTorch_Lightning_2019} interfaces.

\subsection{SC2AnonServerPy}
\label{sec:SC2AnonServerPy}

In the process of extracting the information from the StarCraft~2 replays, the users have the ability to choose if nicknames of the players should be anonymized with a separate tool "SC2AnonServerPy" \cite{Bialecki_2021_SC2AnonServerPy}, this functionality may be key for laboratories that wish to share their datasets with a wider community.

\subsection{Software Functionalities}
\label{sec:SoftwareFunctionalities}

Main functionality of this software collection introduce a repeatable way of working with StarCraft 2 data for research and data analysis. Users need to verify if their specific use case is permitted by the Blizzard End User License Agreement (EULA). Our software package includes file-wrangling tools such as: flattening nested directory structure, data-parallel replay file parsing (extraction), data cleanup, exporting replay data to JSON, and finally data loading into PyTorch \cite{PyTorch2019} and PyTorch Lightning \cite{PyTorch_Lightning_2019}. We have developed a modular system of tools solving specific issues of data processing with expandability in mind.

Main contribution of the work that we present is the ``SC2InfoExtractorGo'' \cite{Bialecki_2021_SC2InfoExtractorGo}, as introduced above. The most important procedure of the data extraction pipeline is showcased in \autoref{fig:file_processing_pseudocode}.

Within ``DatasetPreparator'' \cite{Bialecki_2022_SC2DatasetPreparator} there are multiple scripts that solve specific problems that may be present when researching StarCraft~2, including a wrapper for the "SC2InfoExtractorGo". For example to reproduce the ``SC2ReSet'' and ``SC2EGSet'', scripts from ``DatasetPreparator'' would be executed consecutively as follows:

\begin{enumerate}
 \item ``directory\_flattener.py'' to flatten the nested directory structure of replaypacks that often have a complex structure with meaningful directory naming conventions,
 \item ``directory\_packager.py'' to obtain ``SC2ReSet'' by creating archives of the previously flattened directories,
 \item ``sc2egset\_replaypack\_processor.py'' (requires ``SC2InfoExtractorGo'') to process the replaypacks and obtain the initial version of ``SC2EGSet'',
 \item ``processed\_mapping\_copier.py'' to copy the auxiliary files generated by ``directory\_flattener.py'' to matching output directories.
 \item ``file\_renamer.py'' to rename the files in directories to follow a specific naming convention (e.g., to match the dataset schema),
 \item ``directory\_packager.py'' to obtain the final version of ``SC2EGSet'' by packaging all of the directories in the specified input directory.
\end{enumerate}

\subsection{Code Snippets}
\label{sec:CodeSnippets}

\begin{figure}[H]
    \includegraphics[width=\linewidth]{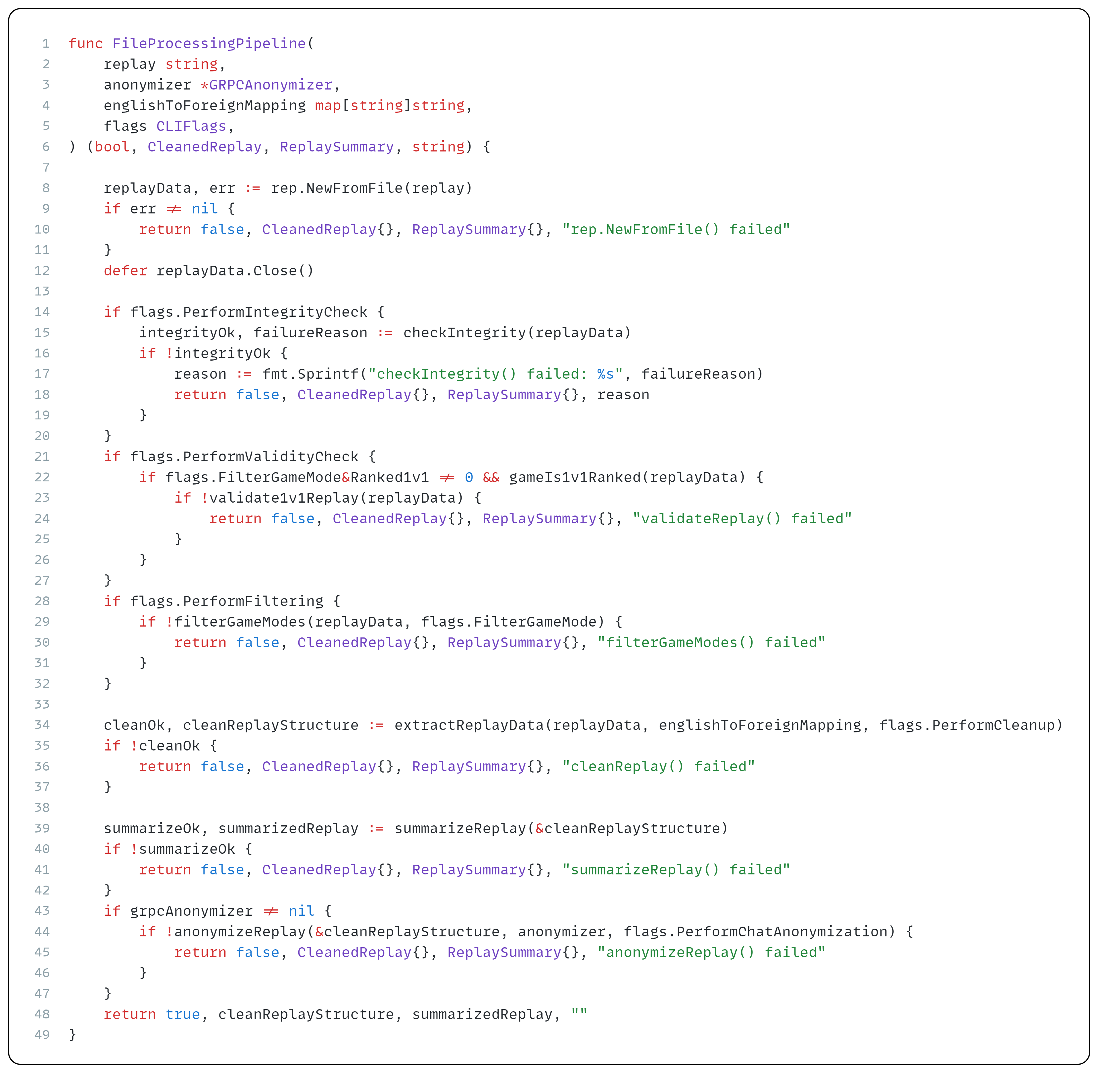}
    \caption{Golang-inspired pseudocode algorithm for processing a single replay file using SC2InfoExtractorGo \cite{Bialecki_2021_SC2InfoExtractorGo}.}
    \label{fig:file_processing_pseudocode}
\end{figure}

Due to the complex nature of our software, and number of operations that are run for every replay, we have created multiple functions that define the steps of the data extraction process with ``SC2InfoExtractorGo''. Function that is ran for every replay is showcased in \autoref{fig:file_processing_pseudocode}.

After the initial processing with a pre-defined pipeline, the output JSON files can be loaded with any programming language capable of reading this format for further processing. In our case this is showcased in ``SC2\_Datasets'' repository, building on top of the JSON files an API for rapid experimentation with ML and AI methods using PyTorch \cite{PyTorch2019} and PyTorch Lightning \cite{PyTorch_Lightning_2019}.

\section{Usage Information}
\label{sec:UsageInformation}

\subsection{DatasetPreparator}

\subsubsection{Usage of Directory Flattener}

It is common for StarCraft~2 tournament replaypacks to be sorted in multiple subdirectories. There is some information to be inferred from the names of the directories. Using this script flattens the diectory structure and prepares it for a simplified further processing.

\subsubsection{Usage of Directory Packager}

Finally, after all of the replaypack, or dataset processing is done, we have prepared a utility script that creates a ``.zip'' archive out of all top-level directories.

\subsubsection{Usage of Processed Mapping Copier and File Renamer}

As described above, after using the ``directory\_flattener.py'' script, one of its side effects is the creation of ``processed\_mapping.json'' file. When preparing a dataset, these files can be treated as additional metadata. Our software in its pipeline includes a script called ``processed\_mapping\_copier.py'', it iterates over each of the input directories and matches it against the directory in the output directory and copies the ``processed\_mapping.json'' file.

To facilitate dataset creation, in most cases replaypack directories are often named after the tournament at which they were collected. File renaming script makes sure that the resulting ``.zip'' archive is renamed to match the tournament name. Some additional auxilliary files are created. This includes: (1) package summaries; containing some basic information about the number of processed replays and other in-game information. (2) previously copied ``processed\_mapping.json'' file considering the directory structure of a replaypack might have been flattened. (3) ``processed\_failed.log'' file containing the information about which files failed to process, and which files were processed successfully. (4) ``main\_log.log'' file, containing all of the logs for debugging.
In case of all of these files the ``file\_renamer.py'' unifies the file names to become prepended with the tournament name e.g., ``main\_log.log'' file becomes ``TournamentName2024\_main\_log.log''.

\subsubsection{Usage of SC2EGSet Dataset Processing Pipeline}

One of the scripts (``sc2egset\_pipeline.py'') simplifies all of the steps required to produce a StarCraft~2 dataset. We use this code to easily reproduce ``SC2ReSet'', and ``SC2EGSet''.

\subsubsection{Usage of SC2EGSet Replaypack Processor}

For a user that wishes to reproduce ``SC2EGSet'' from ``SC2ReSet'', a separate script is available that runs multiple instances of SC2InfoExtractorGo, The script iterates over each of the directory in the input path, and runs the ``SC2InfoExtractorGo'' on each of them with hardcoded parameters. Our solution implements multiprocessing Besides this functionality, the script does not offer much more utility than the original ``SC2InfoExtractorGo'' executable.

\subsubsection{Usage of Other Scripts}

Using the ``sc2reset\_replaypack\_downloader.py'' via its command line arguments makes it is possible to download all available replaypacks published as ``SC2ReSet'' hosted in an Zenodo repository \cite{Bialecki2022ReSetZenodo}. When using ``SC2ReSet'' to run the rest of the processing pipeline, there is no need to execute the ``directory\_flattener.py'', each replaypack was pre-processed before upload. After downloading ``SC2ReSet'' \cite{Bialecki2022ReSetZenodo}, it should be available under the directory as specified by the user.

\subsection{Direct Use of SC2InfoExtractorGo}

Next step pertains to the data extraction with ``SC2InfoExtractorGo''. The input directory for the command line usage of ``SC2InfoExtractorGo'' should reflect the directory where the user stored the replays which they would like to process. To ensure smooth reproduction of SC2EGSet the ``SC2InfoExtractorGo'' should be ran against each of the replaypack directories separately to produce output corresponding to data from a single tournament.

\subsection{Usage of SC2AnonServerPy}

In some cases, the user might want to anonymize the data due to the privacy, ethical, or legislative concerns. We provide additional service named ``SC2AnonServerPy''. As a separate gRPC service it is capable of receiving a string type containing a player nickname, and return a unique identifier as a string type for the requested player. The anonymization server is not constrained to StarCraft~2 data and can be used as long as the user provides the expected input type.

\subsection{Running Experiments With SC2\_Datasets}

\begin{figure}
    \centering
    \includegraphics[width=0.8\linewidth]{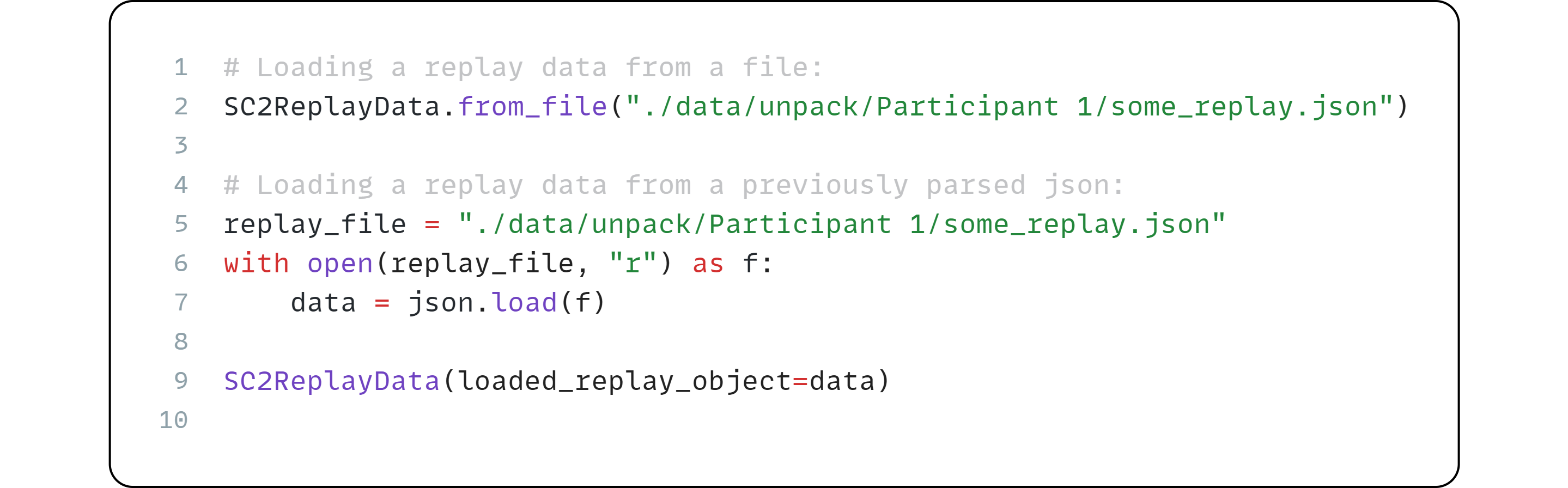}
    \caption{Pictorial representation of code used to load a single replay, as defined in \cite{bialecki_2022_sc2datasets}.}
    \label{fig:single_json}
\end{figure}

\begin{figure}
    \centering
    \includegraphics[width=0.8\linewidth]{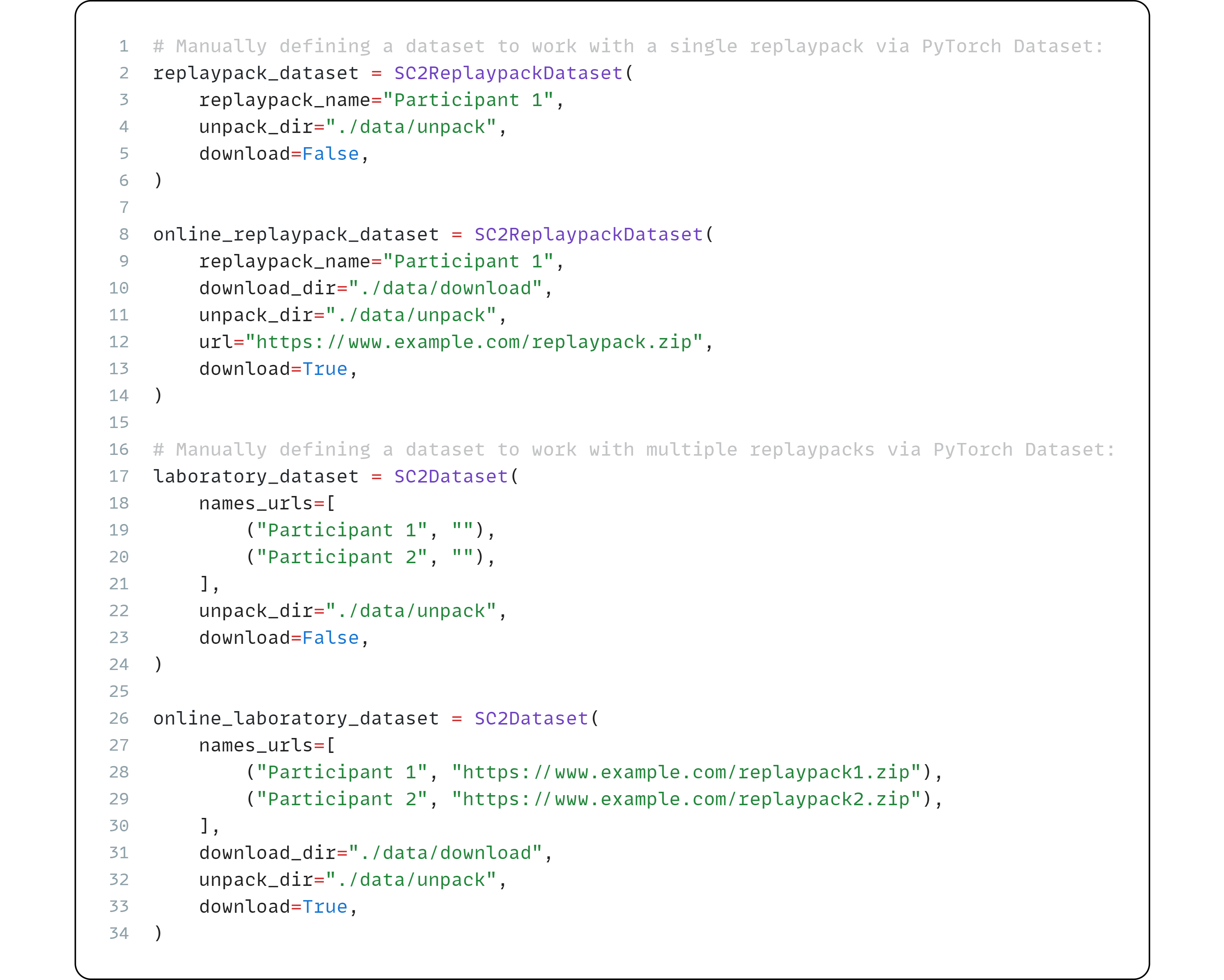}
    \caption{Example usage of the PyTorch \cite{PyTorch2019} dataset interface as defined in \cite{bialecki_2022_sc2datasets}.}
    \label{fig:pytorch_dataset_custom}
\end{figure}

After extracting the data from SC2Replay files, any further processing, and experiments are possible with the ``SC2\_Datasets'' Python package \cite{bialecki_2022_sc2datasets}. Loading a single JSON file following the structure defined in the ``SC2\_Datasets'' parser can be seen on \autoref{fig:single_json}. To load the output of a processed dataset that exists either on the drive or online, the user should initialize a class as visualized on \autoref{fig:pytorch_dataset_custom}. Additionally, PyTorch Lightning \cite{PyTorch_Lightning_2019} datamodule interfaces can be used as they are included in the API Note that the users have full control and customizability of the code. In case of our implementations for ``SC2EGSet'' we provide an interface to use the data. Similar approach can be used with data from other sources, as long as the data is formatted in a way that is compatible with the ``SC2\_Datasets'' parser.

\section{Potential Impact}

There exist many implementations built for the purpose of parsing replay files \cite{ZenodoSC2Reader}. These tools and libraries require expert programming skills to extract and interact with the resulting data. Many research approaches involve scientists that may not posses such expert knowledge in programming, but nonetheless interested in investigating esports (e.g., in psychology, biomechanics, social sciences and humanities -- SSH, and others) \cite{Kegelaers2025,Dupuy2025,Donghee2020}. Lowering the technical overhead needed to interact with in-game data can open gaming and esports to researchers with various non-technical backgrounds. Furthermore, integrating SSH scientists in the research process is not only a requirement in some funding programs, but also a practical necessity, if one aims to conduct socially responsible studies \cite{graf2019bringing,sonetti2020only}.

Before introducing our software, users were bound to write their own tools extracting the data from StarCraft~2 replay files. Our solution outputs easy-to-use JSON files adhering to a specific, well-documented schema definition \href{https://sc2-datasets.readthedocs.io/en/latest/autoapi/index.html}{https://sc2-datasets.readthedocs.io/en/latest/autoapi/index.html}. Additionally, the data extraction toolset efficiently leverages modern multi-core processors (using Golang goroutines), making the process of data extraction faster. This has real implications on day-to-day research, as it allows for faster experimentation and iteration on one's methods.

Within the intended user group, the software was created to assist with the process of StarCraft~2 data processing. Mainly, the software fulfilled the research needs of our team and other collaborating research teams, which led to processing and creating a dataset \cite{Bialecki2023SC2EGSet}. Additionally, an API interface was created to load and work with the data in PyTorch \cite{PyTorch2019} and PyTorch Lightning \cite{PyTorch_Lightning_2019}.

Due to the End User License Agreement (EULA) provisions specified by the game publisher (Blizzard), the commercial use of the extracted game data directly is limited. Nonetheless, one can extract valuable insights from the data and transfer them to the industry in a manner compliant with the EULA. In the past, research conducted on StarCraft~2 data has yielded fruitful ventures in online tooling \cite{URLSc2replaystats,URLSpawningTool,URLSC2Revealed,URLAligulac}; and research \cite{Vinyals2019,Ma2024LLMStarCraft2,Samvelyan2019SMAC,Ferenczi2024sc2_serizlizer}.

\section{Conclusions}
\label{sec:Conclusions}

We conclude that despite there being some software packages available, they often require additional programming skills and knowledge. Our solution provides a simple to use executable file and a set of scripts to work with StarCraft~2 data. Additionally, we conclude that our software solves a very specific infrastructure problem that is prevalent in the gaming and esports research on StarCraft~2.

In its current version our toolset ``SC2Tools'' is capable of simplifying the work associated with handling files used to create a StarCraft~2 dataset. We are planning to keep updating the software to include more tools, features, and functionalities. Additionally, due to the capability of our software to output JSON files, We claim full interoperability with other replay parsing solutions as long as they keep the same output format.

Finally, based on our previous experience in successfully creating a published dataset that was leveraged in other published material \cite{Ferenczi2024sc2_serizlizer}, we conclude that our efforts were not in vain and such infrastructure development may be useful to others.

\section*{Conflict of Interest}
\label{sec:ConflictOfInterest}

We wish to confirm that there are no known conflicts of interest associated with this publication and there has been no significant financial support for this work that could have influenced its outcome.

\section*{Acknowledgements}
\label{sec:Acknowledgements}

We would like to acknowledge various contributions by the members of the technical and research community, with special thanks to: Timo Ewalds (DeepMind, Google), Anthony Martin (Sc2ReplayStats), and András Belicza for assisting with our technical questions.

\section*{Authors' Contributions}
\label{sec:AuthorsContributions}

\begin{itemize}
    \item Conceptualization: Andrzej Białecki;
    \item Methodology: Andrzej Białecki, Piotr Białecki;
    \item Software: Andrzej Białecki;
    \item Technical Oversight: Andrzej Białecki, Piotr Białecki;
    \item Code Review: Andrzej Białecki, Piotr Białecki, Piotr Sowiński;
    \item Writing - Original Draft: Andrzej Białecki;
    \item Writing - Review \& Editing: Andrzej Białecki, Piotr Sowiński, Piotr Białecki, Jan Gajewski;
    \item Licensing and Legal Analysis: Andrzej Białecki, Mateusz Budziak;
    \item Supervision: Andrzej Białecki, Jan Gajewski;
\end{itemize} 

\bibliographystyle{IEEEtran}
\bibliography{sources.bib}

\end{document}